\documentclass[11pt]{iopart}

\usepackage{graphicx}
\usepackage{multicol}

\def\apj{{\it ApJ}}
\def\mnras{{\it MNRAS}}
\def\araa{{\it ARA\&A}}
\def\nat{{\it Nature}}

\begin{document}

\title[Gap-Opening criteria]{Binary-disk interaction:\\ Gap-Opening criteria.}

\author{Luciano del Valle \& Andr\'es Escala }

\address{Departamento de Astronom\'ia, Universidad de Chile, Casilla 36-D, Santiago, Chile}
\ead{ldelvalleb@gmail.com}
\begin{abstract}

We study the interaction of an equal mass binary with an isothermal circumbinary disk motivated by the theoretical and observational evidence of the formation of massive black holes binaries surrounded by gas, after a major merger of gas-rich galaxies. We focus on the torques that the binary produces on the disk and how the exchange of angular momentum can drive the formation of a gap on it. We propose that the angular momentum exchange between the binary and the disk is through the gravitational interaction of the binary and a (tidally formed) global non-axisymmetric perturbation in the disk. Using this interaction, we derive an analytic criterion for the formation of a gap in the disk that can be expressed either via the characteristic velocities of the binary-disk system or in terms of the structural parameters structural parameters $h/a$ and $M(< r)/M_{\rm bin}$. Using SPH simulations we show that the simulations where the binary opens a gap in the disk and the simulations where the disk does not have a gap are distributed in two well separate regions. Our analytic gap-opening criterion predicts a shape of the threshold between this two regions that is consistent with our simulations and the other ones in the literature. We propose an analogy between the regime without (with) a gap in the disk and the Type I (Type II) migration that is observed in simulations of planet-disk interaction (binaries with extreme mass ratios), emphasizing that the interaction that drives the formation of a gap on the disk is different in the regime that we analyze (comparable mass binary).

\end{abstract}

\vspace{2pc}

\maketitle

\section{Introduction}

It's well known that galaxies are systems that can strongly interact gravitationally with each other and even in some cases merge between them. Also, it's widely accepted the fact that galaxies with a significant bulge host a massive black hole (MBH) at its center (Richstone \etal 1998). Therefore, is natural to conclude that these MBHs will interact after a major merger (Milosavljevic \& Merritt 2001, Milosavljevic \& Merritt 2003) and possibly collide. The possible coalescence of MBHs was first considered by Begelman \etal (1980) in a study of the long-term evolution of a black hole binary at the center of a dense stellar system. Initially, dynamical friction brings the two black holes towards the center of the system and the resulting MBH binary continues to shrink via three-body interactions with the surrounding stars. This three-body interactions tends to eject stars from the central region causing the merger eventually to stall (``last parsec''  problem) unless some additional mechanism is able to either extract angular momentum from the MBH binary or refill the stellar ``loss-cone'' (Khan \etal 2011).

There are theoretical (Barnes \& Hernquist 1992, 1996; Mihos \& Hernquist 1996; Barnes 2002; Mayer \etal 2007, 2010) and observational (Sanders \& Mirabel 1996; Downes \& Solomon 1998) evidence that in the mergers of gas-rich galaxies a large amount of gas can reach the central regions of the newly formed system. This gas induces the formation of a MBH binary (Kazantzidis \etal 2005, Mayer \etal 2007, Chapon \etal 2011) and can drive the final coalescence of the binary (Escala \etal 2005; Dotti \etal 2006; Cuadra \etal 2009; Rodig \etal 2011; Escala \& del Valle 2012). 

In all these studies that follow the evolution of a binary embedded in a gas environment is assumed that the gas lies in a disk (circumbinary disk) and the binary has a non-extreme mass ratio ($q \sim 1$). However, they find very different times scales of coalescence for the binary because they explore very different disk to binary mass ratios ($M_{disk}/M_{bin}$). When the mass of the disk is much greater than the binary mass it's found that the coalescence time of the binary is on the order of a few initial orbital times (Escala \etal 2005; Dotti \etal 2006). On the other hand, for disk's with negligible masses, compared to the binary mass, the coalescence time is on the order of several thousand of local orbital times (Artymowicz \& Lubow 1994; Ivanov, Papaloizou, \& Polnarev 1999; Armitage \& Natarajan 2002; Milosavljevic \& Phinney 2005), which  for $\rm M_{BH} \geq 10^{7} M_{\odot}$ is even longer than the Hubble time (Cuadra \etal 2009).

These two regimes of fast (few orbital times) and slow (several thousands of orbital times) migration are also observed in simulations of planets embedded in circumstellar disks (eg. extreme mass ratios; Ward \etal 1989, Ward 1997, Bate \etal 2003, Armitage \& Rice 2005; Baruteau \& Masset 2012, Kocsis \etal 2012a, 2012b). In simulations of planet migration in protoplanetary disks ($q<<1$), these two regimes are defined as Type I and Type II migration; In the Type I regime the perturbation induced by the planet in the circumstellar disk remains small and permits a fast migration of the planet (of the order of a few orbital times) with a characteristic time scale that is inversely proportional to the planet's mass ($t_{migration}\propto M_{\rm p}^{-1}$). When the Hill radius of the planet is greater than the local pressure scale height h ($R_{\rm Hill}>>h$) the perturbation induced by the planet in the disk become important and a gap begins to form, leading to coupled evolution of the planet and the disk on a viscous time scale and therefore, a much longer migration time (Type II migration). In this paper we'll extend the same terminology to the case of  non-extreme mass ratios binaries ($q\sim 1$) interacting with a disk by referring to the fast (slow) coalescence of binary as Type I (II) migration.

 For a comparable mass binary in a disk, as in the star-planet-disk system, the threshold between the Type I and Type II migrations is determined by the formation of a gap in the disk. Our aim in this paper is establish a gap opening criterion that allows us to differentiate these two migrations regimes and determine in which binary-disk systems is expected an efficient coalescence of the MBH binary. For that purpose, we will use a simple model for the binary-disk interaction to derive an analytic expression for the gap-opening criterion. We will test this criterion against a variety of simulations with different model parameters for the binary to disk mass ratio and thickness of the disk.

This paper is organized as follows. After the derivation of an analytic gap-opening criterion in \S 2, we present in \S 3 the numerical method and set of numerical simulations that we use to test our analytic criterion. In \S 4 we describe how we identified the formation of a gap in our simulations and in \S 4.1, we test our analytic gap opening criteria against the numerical simulations. All the results that we find are discussed in \S 5 and in this section are also compare our results with the results of other authors in the literature. Also, in \S 5 we discuss the implications of our results in real astrophysical systems. Finally, our conclusions are presented in \S 6.  

\section{ANALYTIC GAP-OPENING CRITERIA}

The interaction of a binary with a circumbinary disk in cases where the total gas mass is typically much smaller than the mass of the primary, has been widely studied in the context of star/planet formation (Lin \& Papaloizou 1979; Goldreich \& Tremaine 1982; Takeuchi \& Lin, 2002; Armitage \& Rice 2005; Baruteau \& Masset 2012). In these studies is used a linear approximation for the equations of motion of the system and it's found that the interaction between the secondary and the disk is controlled by the sum of the torques arising from the inner and outer Lindblad and corotation resonances. This approach applies to binaries where the primary is much more massive than the secondary ($q=M_{2}/M_{1} \ll 1$) and leads to predictions of the gap structure that are consistent with simulations within the same regime (Ivanov \etal. 1999;Bryden \etal 1999 ;Armitage \& Natarajan 2002; Nelson \& Papaloizou 2003; Haiman \etal. 2009; Baruteau \& Masset 2012). Because of the success of this analysis in the planetary regime ($q\ll 1$), it has been extrapolated to other cases where $q\sim 1$ (Artymowicz \& Lubow 1994, 1996; Gunther \& Kley 2002; MacFadyne \& Milosavljevic 2008) without consider the global non-linear perturbation that is produced by the binary gravitational field in the regime $q\sim 1$ (Shi \etal 2012). This non-linear perturbation breaks the validity of the linearization of the equation of motion.

In this paper we study the binary-disk interaction in the regime $q\sim 1$ without any assumption of linearity. For this purpose we explore if it's possible, due to the tidal nature of the binary-disk interaction, that the gap-opening process can be described by the interaction of the binary with a strong non-axisymmetric perturbation on the disk instead of a resonant process that appears in the linear approximation. This type of interaction between a strong non-axisymmetric perturbation and a binary was investigated in Escala \etal (2004, 2005). They show that, when the gravitational influence spheres (i.e. Hill spheres) of the two component of the binary overlap, a strong tidal non-axisymmetric perturbation is produced in the disk with an ellipsoidal geometry for an equal mass binary ($q=1$). They also found that the symmetry axis of this strong non-axisymmetric perturbation is not coincident with the binary axis but lags behind it, producing a gravitational torque on the binary that is responsible for an angular momentum transport from the binary to the disk.

In this paper we will restrict to the case of an equal mass binary ($q=1$) interacting with a disk, leaving the case $\rm q \sim 1$ but $\rm q \neq 1 $ for a companion paper. In the case $q=1$, is justified to assume that the strong non-axisymmetric perturbation produced in the disk has an ellipsoidal geometry. Therefore, we study the gravitational interaction between the binary and an ellipsoid to derive analytically the exchange of angular momentum between the disk and the binary. This binary-ellipsoid system can be treated as an equivalent one body problem subject to an external gravitational potential caused by the ellipsoidal perturbation. The gravitational potential of a uniform ellipsoid is given by 
\begin{equation}
\Phi(x,y,z)=\pi\, G\, \rho\, (\,\alpha_{0}x^2+\beta_{0} y^2+\gamma_{0} z^2+\chi_{0}\,)\,,
\end{equation}
with $\alpha_0$, $\beta_0$, $\gamma_0$ and $\chi_0$ constant given in Lamb (1879) which depend on the ratio between the principal axes of the ellipsoid. For the symmetry of the ellipsoidal perturbation observed in Escala \etal (2004, 2005) $\gamma_{0}=\beta_{0}$ and $\chi_{0}=0$. Therefore, the Lagrangian of the binary-ellipsoid system in cylindrical coordinates has the form
\begin{eqnarray}
\mathbf{L}&=&\frac{1}{2}\mu\left[ \left(\frac{dr}{dt}\right)^2+ r^2\left(\frac{d\phi}{dt}\right)^2+\left(\frac{dz}{dt}\right)^2 \right]+G\frac{\mu^2}{r}\nonumber\\
&&-\frac{\pi}{2}G\mu\rho(\alpha_0 r^2\cos^2(\Delta \phi)+\beta_0
r^2\sin^2(\Delta \phi)+\beta_0 z^2)\,,
\end{eqnarray}
with $\mu=M_1M_2/M_{\rm bin}$ being the reduced mass of the binary and $\Delta \phi$  the angle between the major axis of the ellipsoid and the binary axis. This angle is approximately constant over the evolution of the system (Escala et al 2004, 2005). From this Lagrangian we derive the Euler-Lagrange equation for the coordinate $\phi$,
\begin{eqnarray}
\frac{d l_{\rm bin}}{dt}&=&\mu\frac{d}{dt}\left(r^2\frac{d\phi}{dt}\right)\nonumber\\
&=&-\pi r^2\left(\frac{M_{\rm bin}}{4}\right)\rho G(\beta_0-\alpha_0)\cos(\Delta \phi)\sin(\Delta \phi)\,,
\label{L}
\end{eqnarray}
where we use $\mu=M_{\rm bin}/4$. This equation expresses the decrease of the angular momentum of the binary $l_{bin}$ and therefore, the injection of angular momentum (torque) from the binary to the disk.

To derive a criterion for the opening of a gap in the disk we compare the gap-opening time scale (determined by the torque that the binary exchange over the disk) with the gap-closing time scales (Goldreich \& Tremaine 1980). The angular momentum $\Delta L$ that must be added to the gas to open a gap of radius $\Delta r$ in a disk with thickness h, is of the order of $\Delta L \approx \rho\,(\Delta r)^2\,h\,r\,v$. The torque that the binary exchange over the disk is $\rm \tau=-dl_{bin}/ dt =r^2\rho\,( M_{\rm bin}/4)\, G\,\pi\,(\beta_0-\alpha_0)\,\cos(\Delta \phi)\sin(\Delta \phi)$. This torque injects an angular momentum $\Delta L$ on a time scale $\Delta t_{\rm open}=\Delta L/ \tau$ and therefore, this is the characteristic time scale to open a gap. This tendency is opposed by viscous diffusion, which fills up a gap of with $\Delta r$ on a timescale $\Delta t_{\rm close}=(\Delta r)^2/\nu$ (Goldreich \& Tremaine 1980) where $\nu$ is the turbulent viscosity of the gas that can be parametrised assuming the standard $\alpha$-prescription $\nu=\alpha_{\rm ss}c_{\rm s}h$ of Shakura \& Sunyaev (1973) where $\alpha_{\rm ss}$ is the dimensionless viscosity parameter and $c_{\rm s}$ is the sound speed of the gas. Gap formation occurs when $ \Delta t_{\rm open} \leq \Delta t_{\rm close}$. We also assume that $c_{\rm s}/v\approx h/r$ with $v$ the circular velocity of the binary-ellipsoid system. Here, based on previous numerical results (Escala et al 2004, 2005), we assume that the binary and the ellipsoid (strong non-axisymmetric perturbation) co-rotate in a circular orbit. If we define $v_{\rm bin}=(GM_{\rm bin}/(8r))^{1/2}$ is straightforward to find that the binary will open a gap in the disk if 
\begin{eqnarray}
\frac{\Delta t_{\rm open}}{\Delta t_{\rm close}}&=&\frac{1}{f}\left(\frac{c_{\rm s}}{v}\right)^3\,\left(\frac{v}{v_{\rm bin}}\right)^2\;\le\; 1 \,,
\label{csv}
\end{eqnarray}
where $f=f(\Delta \phi,\alpha_0 ,\beta_0,\alpha_{\rm ss})=2\pi(\beta_0-\alpha_0)\cos(\Delta\phi)\sin(\Delta\phi)\alpha_{\rm ss}^{-1}$ is a dimensionless function of the geometric parameters of the ellipsoid $(\alpha_0,\beta_0)$, the offset angle $\Delta \phi$ and the viscosity parameter $\alpha_{\rm ss}$. The geometric  parameters of the ellipsoid are approximately constant due to the self-similar behavior of the ellipsoid (Escala et al 2004, 2005). This formulation of the gap opening criterion contains information of the relative weight of the characteristic rotational speed of the binary-disk system ($v$), with the characteristic rotational speed of the isolated binary ($v_{\rm bin}$) and the thermal state of the disk ($c_{\rm s}$). 

Regardless of the exact geometry of the strong non-axisymmetric density perturbation, the gravitational torque from the perturbed background medium onto the binary will have the form $\tau=r^2\rho G\mu K$, where $K$ is a parameter that depends on the geometry of the density perturbation. Therefore, the dimensionless function $f$ has in general the form $f=2K/\alpha_{\rm ss}$ that represents the relative strength between the gravitational and viscous torque.

If we assume that the binary-disk system is in rotational support against the overall gravitational potential on a circular orbit, we can also express this same criterion in terms of the disk structural parameters and the mass of the binary-disk system:
\begin{eqnarray}
\frac{\Delta t_{\rm open}}{\Delta t_{\rm close}}&=&\frac{1}{f}\left(\frac{h}{r}\right)^3\left[1+8\frac{M(<r)}{M_{\rm bin}}\right]\;\le\;1\,,
\label{hr}
\end{eqnarray}
where $M(<r)$ is the total mass enclosed by the orbit of the binary.

\section{INITIAL CONDITIONS AND NUMERICAL METHOD}

In this section we present simulations that follow the evolution of a binary embedded in a gaseous disk. We use a natural units system in our simulations: [mass]=1, [distance]=1 and we set the gravitational constant G=1. In our internal units, the initial radius of the disk is $R_{\rm disk}=30$, and the total gas mass is $M_{\rm disk}=33$. The density of the disk is given by

\begin{eqnarray}
\rho(R,z)&=&\frac{\Sigma_{0}}{2\, h_{\rm c}}\, \frac{R_0}{\, R_{\rm c}} \cosh^{-2}\left(\frac{z}{h_{\rm c}}\right)\qquad\qquad  R\leq R_c\\  
         &=&\frac{\Sigma_{0}}{2\,h_{\rm c}}\,\frac{R_0\,R_c}{R^2} \cosh^{-2}\left(\frac{z}{h_{\rm c}}\frac{R_{\rm c}}{R}\right)\qquad R_c\le R \leq R_{\rm disk} \,,
\end{eqnarray}

where $R_{\rm c}=3$, and $h_{\rm c}$ are the radius and thickness of the central zone of the disk where the surface density is constant. With this density we obtain a surface density $\Sigma(r)=constant$ for $R<R_{\rm c}$ and $\Sigma(r)\propto R^{-1}$ if $R>R_{\rm c}$.  The vertical distribution of the disk changes over the evolution of the system, but the initial vertical distribution chosen at least prevents an initial vertical collapse on the disk. The binary system have an initial circular orbit of radius $r_0$, coplanar and co-rotating with the disk, a total mass $M_{\rm bin}$, and a mass ratio between the two components of the binary $q=M_2/M_1=1$. We explore the parameter space $r_0\in[0.5,4]$, $M_{\rm bin}\in[1,33]$, and $h_{\rm c}\in[0$.$8,3]$ with 25 simulations (see table 1).

In addition to the disk and the binary we include a fixed Plummer potential (Plummer 1911) that helps to stabilize the disk and also, when we apply our result on the study of SMBH binaries in section \S 5, will mimic the existence of an external stellar component. The total mass of the Plummer potential is $M_{\rm p}=6$.$6$ ( $\sim\%12$ of the total mass of the disk) and its scale radius is $R_{\rm p}=19$.$13$. The applicability of the gap-opening criterion derived in \S 2, remains valid even in the presence of this Plummer potential because its mass can be included in the total mass enclosed by the orbit of the binary $M(<r)$.

We model the gaseous disk with a collection of $2\times10^5$ SPH particles of gravitational softening 0.1. We use a stable ($\rm Q\ge 1$) isothermal disk to avoid fragmentation, to simplify the testing of our analytic criterion derived in the section \S 2. In table 1 we tabulate the minimum value of the parameter Q of Toomre of each simulation. Since the total potential of the system  (Binary-Disk-Plummer) is non-axisymmetric and therefore lacks of a well defined circular velocity, we assume initially a symmetric representation of the binary potential to compute the circular velocity of the disk (see Appendix A for details). For the binary we use 2 collisionless particles with gravitational softening of 0.1. We run each simulation for 10-15 binary orbital times using the SPH code called GADGET2 (Springel 2005, Springel \etal 2001). 

In addition, to test the numerical convergence of the results, we run simulations with 1 million SPH particles and find that the conclusions found in \S 4 for our analytic gap-opening criterion do not change (Appendix B). 

{\small
\begin {center}
\centerline{Table 1: Run Parameters}
\vspace{0.4 cm} 
\begin{tabular}{ccccc} \hline 

RUN  & $\rm r_{0}/R_{\rm disk}$ & $\rm M(<r)/M_{\rm bin}$ &  $\rm (c_{\rm s}/v_{\rm bin})^2$ & $\rm Q_{\rm min}$  \\ \hline \hline 
A1  & 0.133 &0.05 &  0.088 & 2.62 \\
B1  & 0.1   &0.06  &  0.080 & 2.50 \\  
B2  & 0.1   &0.06  &  0.159 & 2.50 \\  
B3  & 0.1   &0.06  &  0.199 & 2.49 \\  
B4  & 0.1   &0.06  &  0.279 & 2.48 \\  
B5  & 0.067 &0.06  &  0.053 & 2.50 \\  
C1  & 0.1   &0.1  &  0.133 & 2.22\\
C2  & 0.067 &0.1  &  0.089 & 2.20 \\
C3  & 0.1   &0.1  &  0.199 & 2.20\\
C4  & 0.067 &0.1  &  0.178 & 2.20\\
C5  & 0.067 &0.1  &  0.222 & 2.20\\
C6  & 0.1   &0.1  &  0.465 & 2.19\\
C7  & 0.067 &0.1  &  0.311 & 2.19\\
D1  & 0.1   &0.12 &  0.159 & 2.14\\
D2  & 0.05  &0.12 &  0.080 & 2.15\\
D3  & 0.05  &0.12 &  0.199 & 2.14\\
E1  & 0.067 &0.2  &  0.178 & 2.01\\ 
E2  & 0.033 &0.2  &  0.088 & 2.01\\ 
E3  & 0.033 &0.2  &  0.219 & 1.99\\ 
E4  & 0.067 &0.2  &  0.623 & 1.98\\ 
F1  & 0.017 &0.4  &  0.090 & 1.85\\
F2  & 0.033 &0.4  &  0.175 & 1.85\\
F3  & 0.067 &0.4  &  0.356 & 1.85\\
G1  & 0.017 &2.0  &  0.452 & 1.70\\
G2  & 0.067 &2.0  &  1.779 & 1.70 \\
\hline 

\label{TABLA}
\end{tabular}
\end{center}
}
\section{RESULTS}

In order to test our analytic gap opening criterion against SPH simulations, we first need a criterion that determines numerically which of our simulations opens a gap in the disk and which ones don't. To determine whether a gap is opened or not, we explore the radial density profile of each simulation and the evolution of the binary separation.

\begin{figure}[h!]
\centering 
\includegraphics[height=5.5cm]{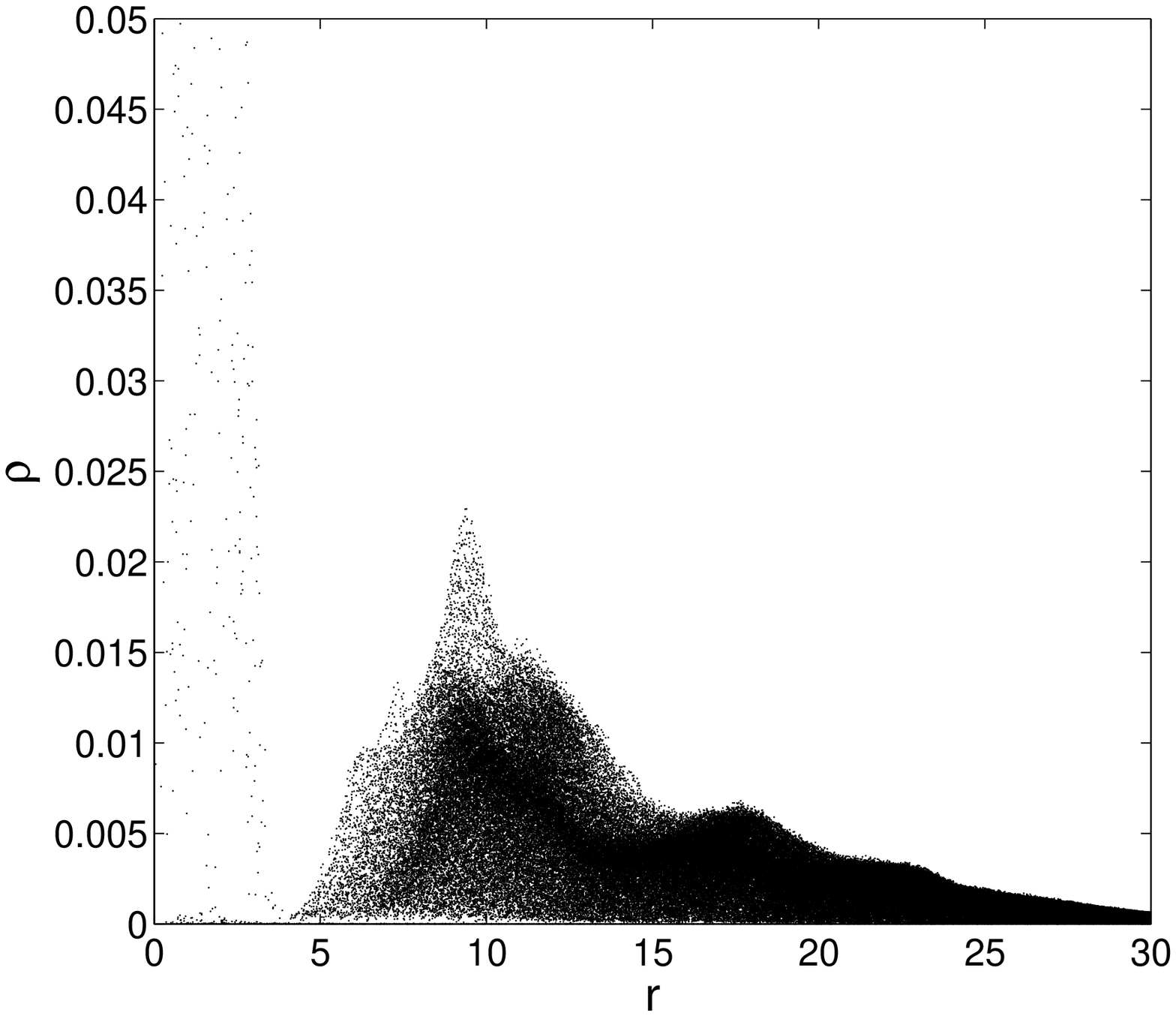}\includegraphics[height=5.5cm]{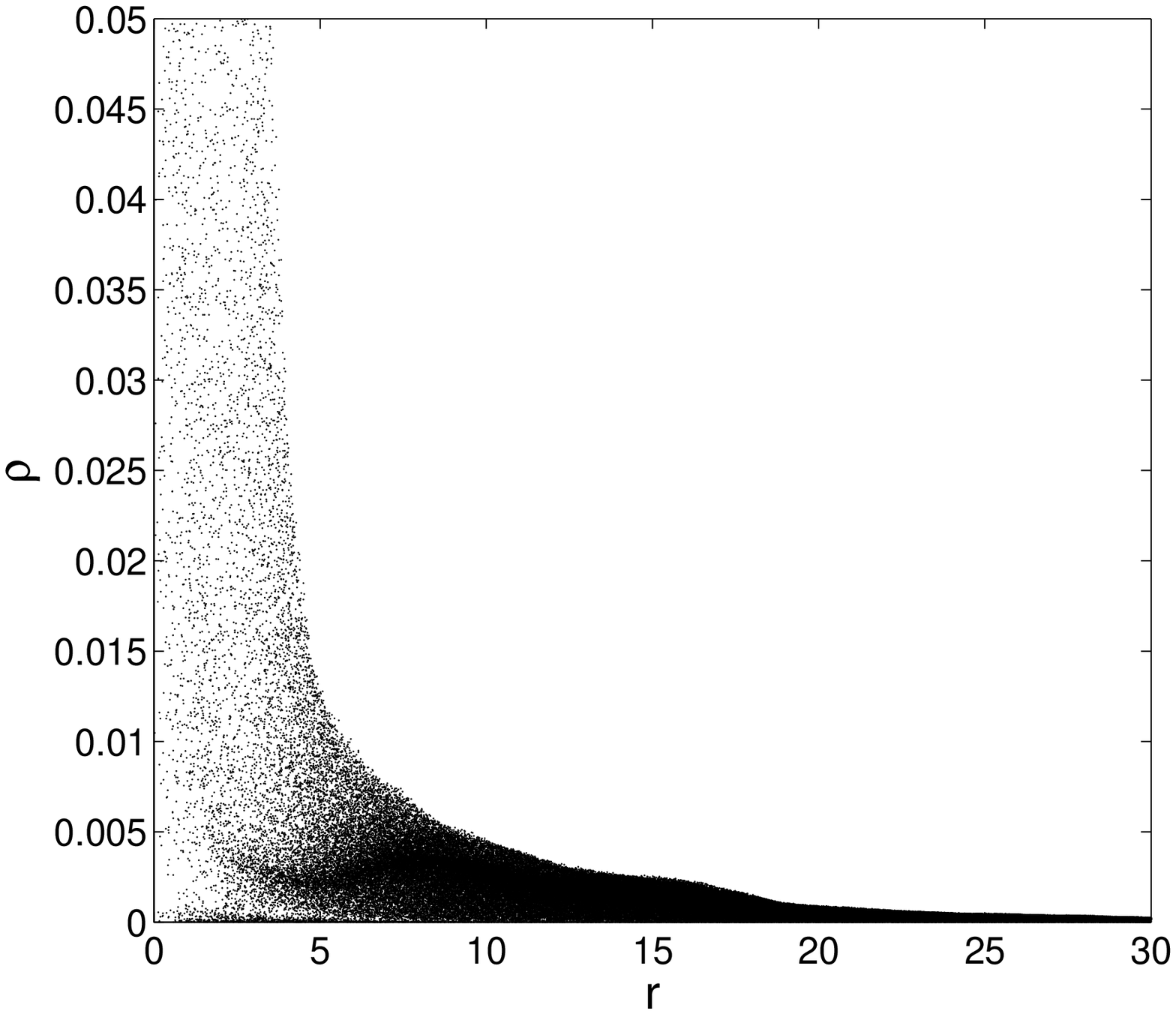}\\
\hspace{0.5cm}\includegraphics[height=4.5cm]{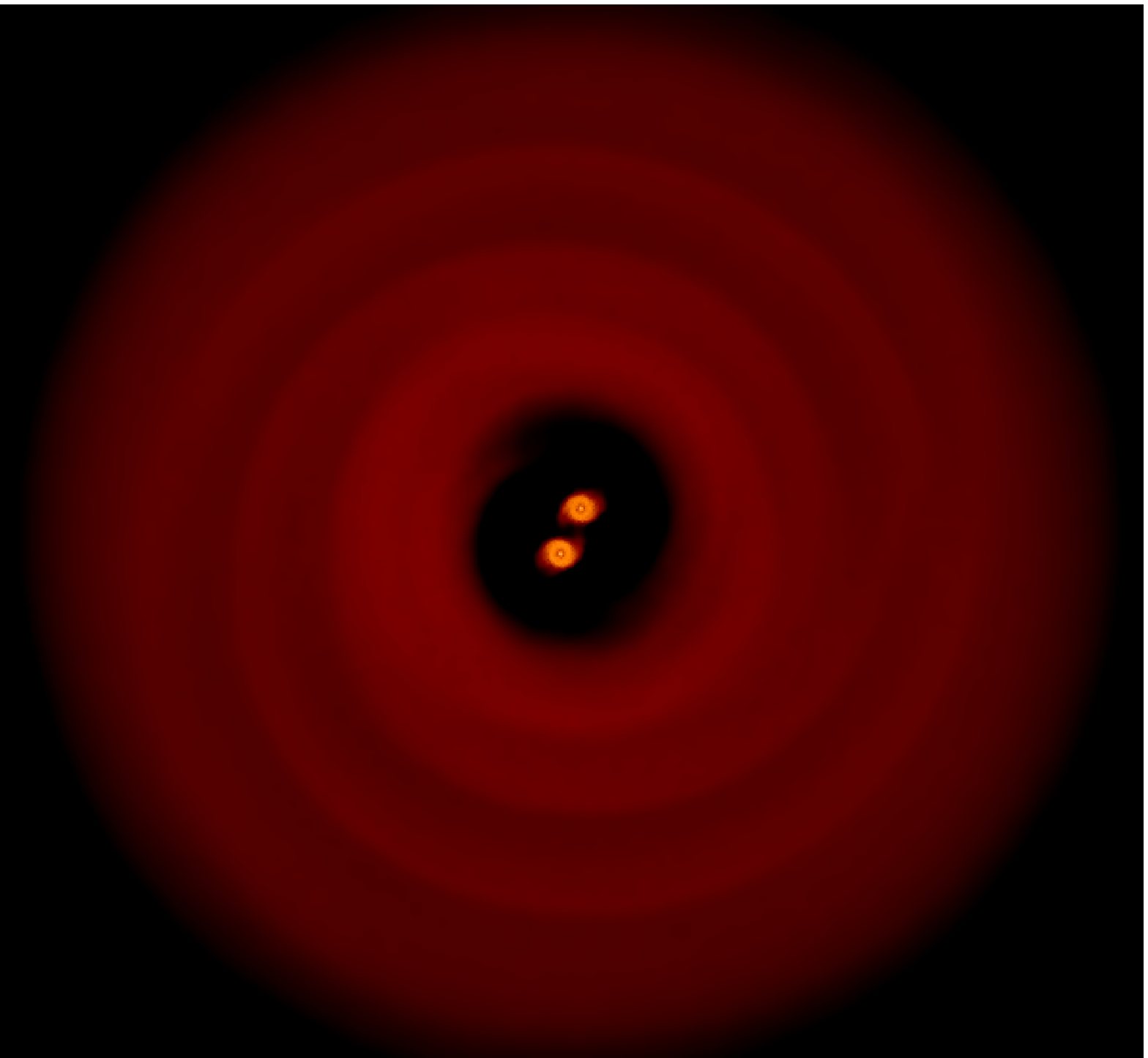}\hspace{1.4cm}\includegraphics[height=4.5cm]{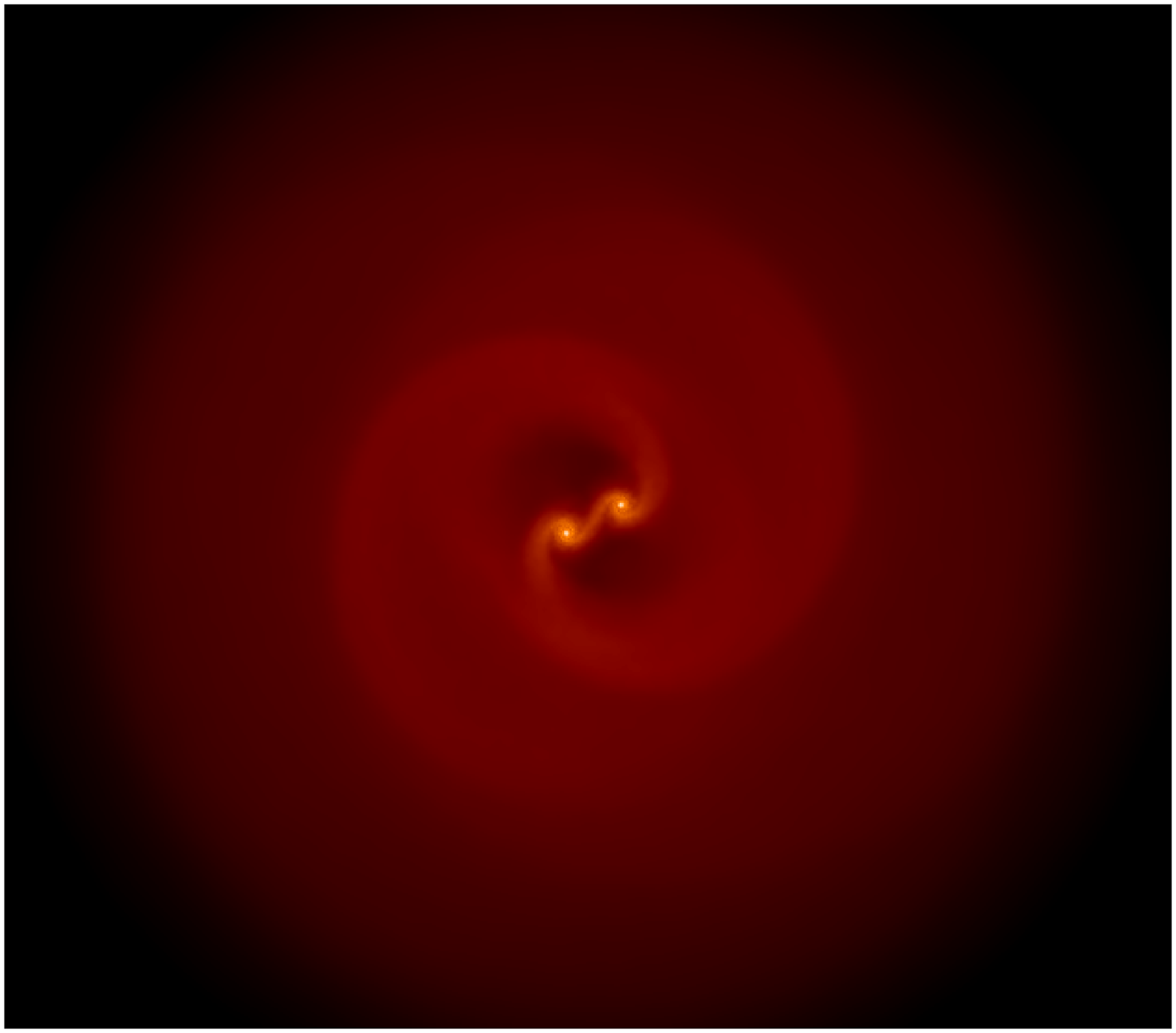}
\caption{In the top: the density projection in the plane x-y of the simulation C2 (left) and F3 (right). In the bottom: The density profiles of the simulation C2 (left) and F3 (right). In this set of graphics is shown the correlation between the  simulations with a gap (without a gap) and a density profile with a pile up of gas whose maximum value is $\rho_{\rm peak}>0.015$ ($\rho_{\rm peak}<0.015$) (see section \S 4)}
\label{open-close}
\end{figure}

If the binary opens a gap in the disk, it must have a flow of gas from the central to the outer regions of the disk. This flow produces a ``pile-up'' of gas on the perimeter of the gap. This type of pile-up is also expected to form on the evolution of a circumbinary disk that interacts with an unequal mass binary but in this regime ($q\ll 1$) is associated to a drop in the density of the disk and not necessarily to the formation of a gap (Kocsis \etal 2012). The pile-up formed in the evolution of our simulations ($q=1$) is represented by a peak in the density profile and its maximum is correlated to the existence of a gap in the disk. In the figure \ref{open-close} we show that the density plot has a peak or pile-up if the disk has a gap. For the cases that the disk doesn't have a gap, this pile-up is less prominent or doesn't exist.

 We also explore if the gap can be identified as a drop in the azimuthal mean density. Although in some cases the existence of a gap is correlated with a drop in the mean azimuthal density, in many other cases there are streams of gas or inner disks around individual components of the binary that hide the existence of a gap in an azimuthally averaged density plot. Therefore, we decided to use the peak in the density profile as our first indicator for the existence of a gap. In addition to this, we follow the evolution of the binary separation. This separations tell us whether the binary is migrating or not to the inner regions of the disk on a time scale comparable to its orbital time $t_{\rm orb}$. In order to the binary migrates is necessary an efficient angular momentum transfer that cannot be reached if the material that participates in this angular momentum exchange is diffused to the outer regions as in the open-gap case. Therefore, a binary that is migrating to the center of the disk on a time scale $\rm \sim\, t_{\rm orb}$, is only consistent  with a disk that does not have a gap (Type I migration).

 Considering the density profile and the possible migration of the binary to the inner region of the disk, we called {\it opened} to a simulation in a given time $t$, if it has a peak or pill-up in density with $\rho_{\rm peak}\geq 0.015$ (in internal units) and if the binary separation do not decreases by more than 10 \% in a orbit. If a simulation in a time $t$ have a maximum density of the pile up $\rho_{\rm peak}\le 0.015$ and the binary separation decreases more than 10 \% in a orbit, we called {\it closed}. Otherwise we called it {\it indefinite}.

We analyze our simulations at the times $t$ in which the binary complete 2, 3, 5, 7, 10 and 15 orbits. For all the simulations in each of these time intervals, we define it as {\it opened}, {\it closed} or {\it indefinite} if the gap in the disk is open, close or indefinite respectively. 

\subsection{Testing the Gap-opening Criterion}

In order to test our analytic gap-opening criterion we plot the velocity parameters of the system $(\;(\,v_{\rm bin}\,/\,v\,)^2\,,\,(\,c_{\rm s}\,/\,v\,)^3\;)$. This parameters are present in one of the two equivalents gap-opening criteria (equation \ref{csv}). In a first attempt we try to determine the values of the parameters $(\,1+M(<r)\,/\,M_{\rm bin}\,)^{-1}$ and $\,(h\,/\,r\,)^3$ that are present in the other gap-opening criterion (equation \ref{hr}) but were poorly defined because the structural parameters of the disk (thick, radius and density profile) can vary strongly (variations on the order of 30-70\% ) in the vicinity of the binary, in time-scales shorter than the  orbital time-scale of the binary, due to the local action of the gravitational potential of the components of the binary.

The velocity parameters are derived from the dynamical state of the binary that, unlike the structural parameters of the disk, have variations in times scales greater or at least comparable, to the time scale of the orbit of the binary. Therefore, to compute the structural parameters of the system we derive them from the velocity parameters. These structural and velocity parameters are related through the relations $8(M(<r)/M_{\rm bin})+1=(v/v_{\rm bin})^2$ and $(h/r)^3=(c_{\rm s}/v)^3$.
 
\begin{figure}[h!] 
\centering
\includegraphics[height=7cm]{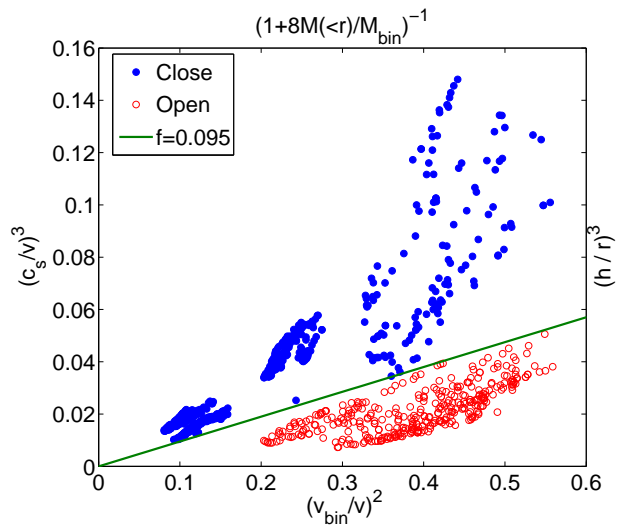}
\caption{The figure shows the cubic ratio between the sound speed of the gas and the rotational velocity of the binary-disk system $(c_{\rm s}/v)^3$ plotted against the quadratic ratio between the rotational velocity of the isolated binary and the rotational velocity of the binary-disk system $(v_{\rm bin}/v)^2$. As $(v_{\rm bin}/v)^2=(1+8M(<r)/M_{\rm bin})^{-1}$ and $(c_{\rm s}/v)^3=(h/r)^3$ we also include this axis labels where $M(<r)/M_{\rm bin}$ is the enclosed to binary mass ration and  $h/r$ is the ratio between the disk thickness and half the binary separation ($r=a/2$). The red circles are simulations where the binary has opened a gap in the disk ({\it opened} simulations) and the blue filled circles are simulations where the disk does not have a gap ({\it closed} simulations) (see section \S 4). The green line is the threshold between the {\it opened} simulations and the {\it closed} simulations that is predicted by our analytic gap-opening criterion. Below the green line are the {\it opened} simulations and above the line are the {\it closed} simulations. The slope of the interface is the function $f(\Delta \phi,\alpha_0 ,\beta_0,\alpha_{\rm ss})$.}
\label{param_VelDis}
\end{figure}

In figure \ref{param_VelDis}, we plot two groups of simulations; the {\it opened} and the {\it closed} simulations according to our definition. The {\it opened} simulations are represented by red open circles and the {\it closed} simulations are represented by blue filled circles. We find that the {\it opened} and {\it closed} simulations are distributed in two different regions. The interface that separates the set of simulations that are {\it opened} from the set of simulations that are {\it closed} can be modeled by a straight line (which slope is f as we will show in the next section). The linear nature of the interface indicates that the functional dependence of our gap-opening criterion on the parameters of the disk-binary system works fine, in the range of parameters that we explore. This encourages us to explore in more detail the significance of this linear interface.

An important consequent of Figure 2 is that the extension of the standard gap-opening criterion derived for the planetary regime ($q\ll 1$) to the regime $q\sim 1$ doesn't predicts the right shape of the interface line between the {\it opened} and {\it closed} simulations. For example the standard gap opening criterion in Lin \& Papaloizou (1986) (Equation (4)) for the case $q = 1$ can be expressed as $(h/r)^3\le (1/40\alpha_{\rm ss} )^{3/5}$ which corresponds to an horizontal line in Figure 2, which clearly cannot explain the distribution of {\it opened} and {\it closed} simulations in  \ref{param_VelDis}.\\

\subsection{Determining an average value for $f(\Delta \phi,\alpha_0,\beta_0,\alpha_{\rm ss})$}

 The interface-line that was shown in the previous section separates the parameter space, between the {\it opened} and {\it closed} simulations and can be interpreted as the critical case in which a simulation have equals opening and closing times (i.e $\Delta t_{open}=\Delta t_{close}$). The slope of the interface-line that separates the set of  {\it opened} simulations from the set of {\it closed} simulations in the velocity variables graph (figure \ref{param_VelDis}) is 
\begin{eqnarray}
m=\left(\frac{v}{v_{\rm bin}}\right)^2\,\left(\frac{c_{\rm s}}{v}\right)^3=\left(\frac{h}{r}\right)^3\left[1+8\frac{M(<r)}{M_{\rm bin}}\right]\,.
\label{slope}
\end{eqnarray} 
We find a value for the slope of the interface-line approximately of $m=0.095$. 

From the equation \ref{csv} for the case $\Delta t_{\rm open}=\Delta t_{\rm close}$ we can estimate that the dimensionless function $f$ that contains the information of the relative strength between the gravitational and viscous torques 
, is on average $f(\Delta \phi,\alpha_0 ,\beta_0,\alpha_{\rm ss})=m$. Replacing this numerical value for $f$ in the equations \ref{csv} and \ref{hr} we can express the gap-opening criterion for an equal mass binary as 

\begin{eqnarray}
\,\left(\frac{v}{v_{bin}}\right)^2\left(\frac{c_{\rm s}}{v}\right)^3\;\le\; 0.095\,.
\label{hr2}
\end{eqnarray}
or
\begin{eqnarray}
\,\left(\frac{h}{r}\right)^3\left[1+8\frac{M(<r)}{M_{\rm bin}}\right]\;\le\; 0.095
\label{csv2}
\end{eqnarray}

As $f\propto \alpha_{\rm ss}^{-1}$ any changes on the viscosity parameter $\alpha_{\rm ss}$ will change the slope of the interface-line of the figure \ref{param_VelDis}. If we increase the value of $\alpha_{\rm ss}$ the slope will be less steep and the number of {\it closed} simulations will increase. This is consistent with the fact that with an increase of the viscosity it will be harder for the binary to open a gap on disk. In our simulations we can estimate $\alpha_{\rm ss}$ from the value of the SPH parameter of artificial viscosity $\alpha_{\rm sph}$ (Artymowicz \& Lubow 1994, Murray 1996, Lodato \& Price 2010, Taylor \& Miller 2011). The value that we estimate ranges between $\alpha_{\rm ss}\approx 0.008 \sim 0.016$. However, independent to the  exact value; the functional dependence of our gap-opening criterion remains unchanged.

\subsection{Transition from {\it closed} regime to {\it opened} regime}

Only one of our 25 simulations (simulation C5 in table \ref{TABLA}) evolves from a state where the disk has no gap ({\it closed}) to a state where the disk has a gap ({\it opened}). The binary in this simulation initially migrates to the center of the disk driven by the action of the tidal torque. As the binary migrates to the center, eventually reaches a separation where the torque exerted into the disk is enough to open a gap. It is important to emphasize that this is only a particular case that already start near the {\it opened} region and doesn't mean that any binary as migrates to the center will be able to open a gap. Moreover, in most cases studied, we have binaries that as migrate to the center never fulfil the condition to open a gap, because the mass ratio $M_{\rm bin}/M(<r)$ doesn't grows fast enough compared to the growth of the thickness $h/r$.
\begin{figure}[h!] 
\centering
\includegraphics[height=7cm]{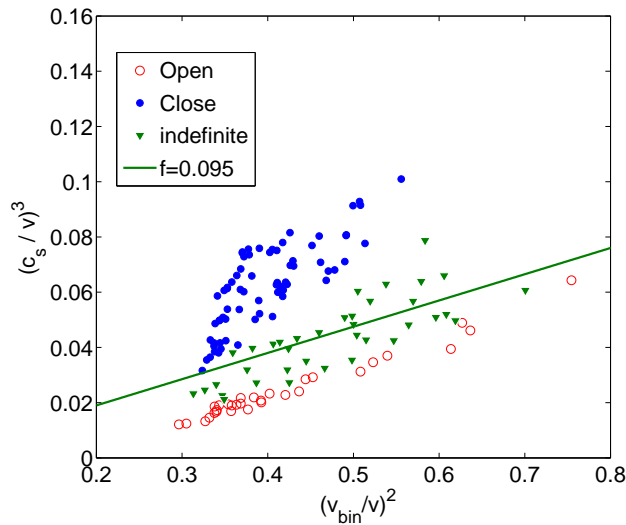}
\caption{The figure shows the cubic ratio between the sound speed of the gas and the rotational velocity of the binary-disk system $(c_{\rm s}/v)^3$ plotted against the quadratic ratio between the rotational velocity of the isolated binary and the rotational velocity of the binary-disk system $(v_{\rm bin}/v)^2$. We plot three groups of points representing three groups of different regimes for a same simulation in different times (C5). The three regimes are {\it opened} (red open circles), {\it closed} (blue filled circles) and a transitory {\it indefinite} regime (green triangles) according to our definition (see section \S 4). The green line is the interface between the {\it opened} population of parameters (below the line) and the {\it closed} population of parameters (above the line). The slope of the interface is the function $f(\Delta \phi,\alpha_0 ,\beta_0,\alpha_{\rm ss})$. The three regimes are distributed in the plot consistently with the criterion.}
\label{Trans}
\end{figure}
 It's interesting to visualize the transition of the simulation C5 between the {\it opened} and {\it closed} gap regions in the velocity parameters space $((v_{\rm bin}/v)^2,(c_{\rm s}/v)^3)$ (figure \ref{Trans}). In this plot we find that the three regimes of evolution in this simulation are distributed consistently with our criterion derived in \S 2, with the  {\it closed} points above the threshold line ($f=0.095$), the {\it opened} points below the threshold  and the {\it indefinite} ones, oscillating around it.
\newpage
\section{DISCUSSION}

In our set of simulations we identify two regimes of binary-disk interactions: with a gap ({\it opened}) and without a gap ({\it closed}). We found that in general less massive and thinner disks are strongly perturbed by the binary and a gap is formed in them. On the other hand, an increase of the mass or thickness of the disk tends to close this gap and in some cases, even preclude its formation. We also found that the possibility of formation of such a gap is more sensitive to the thickness than to the mass of the disk. This is in agreement with our analytic criterion that has a linear dependence in the mass ratio $M_{\rm bin}/M(<r)$ and a cubic dependence in the ratio between the pressure scale height of the disk and the binary separation ($h/a$).

 We find good agreement between our test-simulations and our analytic criterion, this support the role of the interaction between the binary and a strong non-axisymmetric perturbation as the responsible of the formation of the gap in the disk, instead of a resonant process. However, the resonances that operates in regions far from the binary where the linearization of it gravitational potential is a good approximation ($r>>a$), play an important role in the later evolution of the gap, ``clearing'' it edges and driven the angular momentum transport from the binary to the disk when the gap is already formed (Artymowicz \& Lubow 1994; MacFadyen \& Milosavljevic 2008; Cuadra \etal 2009; Roedig \etal 2012).

\begin{figure}[h!] 
\centering
\includegraphics[height=7cm]{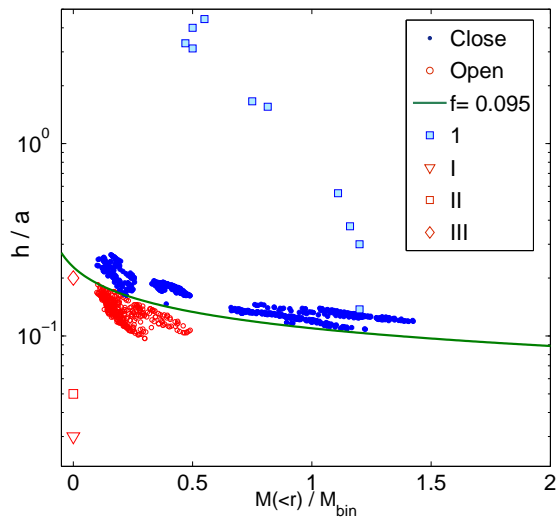}
\caption{The ratio between disk thickness and binary separation $(h/a)$ plotted against the enclosed to binary mass ratio ($M(<r)/M_{bin}$), for the simulations with a gap in the disk (open red circles) and without gap in the disk (blue filled circles). The continuous green curve represent the interface of this two populations of parameters (f = 0.095). The other points in the figure, are simulations already in the literature: Escala \etal 2005 (1), Artymowicz \& Lubow 1994 (I), Artymowicz \& Lubow 1994 and MacFadyen \& Milosavljevic 2008 (II), Cuadra \etal 2009 (III). The points with arabic number labels, are simulations where the disk don't have a gap and the points with roman number labels are simulations where the disk have a gap. All the points are consistent with our analytic gap-opening criterion. The qualitative behavior predicted by the opening criterion is reproduced in this graph, as  $h/a$  decreases, the binary-disk interaction tends to open a gap. Also if we decreases $M(<r)/M_{bin}$,  the interaction tends to open a gap again.}\label{PARAM_Num}
\end{figure}

 To improve the robustness of our study in figure \ref{PARAM_Num} we plot the parameters $\:M(<r)\,/\,M_{\rm bin}\:$ and  $\:h\,/\,r\:$ (being $r=a/2$ with $a$ the binary separation) for our simulations and the {\it  close} and {\it opened} simulations from the literature (Artymowicz  \& Lubow 1994; Artymowicz \& Lubow 1996; Escala \etal 2005; MacFadyen \& Milosavljevic 2008; Cuadra \etal 2009). In this figure we also plot the interface-curve ($\Delta t_{\rm open}=\Delta t_{\rm close}$) that for this space of parameters has an inverse-cubic form. For the cases of simulations with non-self gravitating disks (Artymowicz \& Lubow 1994; Artymowicz \& Lubow 1996; MacFadyen \& Milosavljevic 2008) we assume $M(<r)/M_{\rm bin}= 0$. All the {\it opened} simulations that we find in the literature begins with a gap in the disk ($M(<r_0)/M_{\rm bin}=0$) or the disk have a low mass compare to the binary mass, but we dont find any simulation where the formation of a gap is study for a massive disks ($M(<r_0)/M_{\rm bin}>1$). In the figure \ref{PARAM_Num} we can see that the simulations in which the disk begins with a gap, or it's open in the evolution of the simulation, (red open points I, II and III) are below the interface curve, consistent with the gap-opening criterion. The {\it closed} simulations (squares in figure \ref{PARAM_Num}) represent the final evolution of simulation that begins with initial condition as the used by Escala \etal 2005 and Dotti \etal 2006. These simulations are above the interface line and are also consistent with the gap opening criterion. Therefore all the simulations from papers that we refer from the literature are consistent with the gap-opening criterion independently of the different equations of state that are used for the disks. 

In the final evolution of the simulations from Escala \etal (2005), squares in figure \ref{PARAM_Num}, is found that although the binary shrinks their separation up to distances where its gravitational potential begins to dominates locally and the enclosed mass is comparable or smaller than the binary mass ($\rm M(<r)/M_{bin}\leq 1$), the binary fails to form a gap in the disk. This is because the scale height of the disk remains roughly constant as their separation $a$ (equals to $2r$) decreases, resulting in an increases of $h/a$ that compensates the decrease in $\rm M(<r)/M_{bin}$ in such a way that the system always lies above the threshold line $f=0.095$. 

We expect that a situation like the one found in  Escala \etal (2005), should happen in galaxies with a gas rich nucleus, such as nuclear disks in ULIRGs (Downes \& Solomon 1998), Submillimiter Galaxies (Chapman \etal 2003; Chapman \etal 2005; Tacconi \etal 2006; Swinbank \etal 2010) and in general, on protogalaxies at the early universe. This is because both observations (Downes \& Solomon 1998; Genzel \etal 1998) and simulations (Mayer \etal 2010; Bournaud \etal 2011) found typical  disk's scale height larger than several tenths of parsecs and gaseous masses larger than $\rm 10^{9} \, M_{\odot}$ (more  than the mass of  most massive MBHs binaries). Moreover,  we determine f in our gap opening criteria (equation \ref{csv}) with simulations using an $\alpha_{ss} \approx 0.01$ ($\rm f \propto \alpha_{ss}^{-1}$) and  gas in massive nuclear disks is expected to be globally unstable. In such a case, the torques are significantly larger, with $\alpha_{ss}$ of order unity (Krumholz \etal 2007; Escala 2007). Therefore, in this situation we expect that to open a gap will be even harder than what we found in our simulations.

On the other hand, in mergers of gas poor galaxies we expect MBHs binaries with less massive and thinner circum-binary disks in which a gap will be easily opened. In such cases, the MBH-disk interaction will be better described by Type II simulations (Artymowicz \& Lubow 1994; Artymowicz \& Lubow 1996; MacFadyen \& Milosavljevic 2008; Cuadra \etal 2009). However, in order to make concrete predictions, is needed high-resolution galaxy mergers simulations that includes feedback processes from star formation and BH accretion, to study under which conditions the criteria given by equation \ref{csv} is fulfilled or not and therefore, the binary-disk interaction will be Type I or II.

Our study has also implications for the formation of gaps around binary protostars, in which both Type I and II interactions can be present at different epochs of their formation. In simulations of early epochs in the formation of binary stars, for example Boss (1982), Bate \& Bonnell (1997) that studies the evolution of binary proto-stellar seeds within a collapsing cloud, is found that the  interaction of the binary with the ambient gas is through a non-axisymmetric overdensity like the one studied in this paper as Type I. On the other hand, in studies of the final stages in the formation of binary stars, simulations of proto-planetary disks around binary stars recurrently found gaps that produce disk truncation at $\rm r_{\rm t} \sim 3/2 a$  (Artymowicz \& Lubow 1994; Artymowicz \& Lubow 1996; Gunther \& Kley 2002). Between both limiting epochs in the formation of binary stars, it must be a transition epoch that is given by the threshold condition studied in this paper (equation \ref{csv}). When this transition occurs in binary formation, there must be implications in the final separations and masses of the binaries, since it corresponds to a dramatic transition in both the migration and accretion timescales.  

\section{CONCLUSIONS}

We study the interaction of an equal mass binaries with an isothermal gas disk. We focus on the torques that the binary produces on the disk and how this exchange of angular momentum can drive the formation of a gap on it. 

We propose that the angular momentum exchange between the binary and the disk is through the gravitational interaction of the binary and a tidally formed strong non-axisymmetric perturbation in the disk. From this gravitational torque we derive an analytic criterion to determine if the binary will open a gap or not in the disk. 

The analytic gap opening criterion that we derive depends on two dimensionless parameters  $(v_{\rm bin}/v)^2$ and $(c_{\rm s}/v)^3$ where $v$ is the rotational velocity of the binary-disk system, $v_{bin}$ is the rotational velocity of the isolated binary and $c_{\rm s}$ is the sound speed of the disk. We also show that, through a transformation of variables, the criterion can be rewritten into other two parameters; the first is a function of the enclosed to binary mass ratio $(1+M(<r)/M_{\rm bin})$ and the second is the cubic ratio between the disk thickness and the binary separation $(h/r)^3$. 

 Using SPH simulations we show that the simulation where the binary opened a gap in the disk ({\it opened} simulations), and the simulations where the disk does not have a gap ({\it closed} simulations), are distributed in two well separate regions in the parameter spaces $(\:(\,v_{\rm bin}\,/\,v)^2\, , \,(\,c_{\rm s}\,/\,v\,)^3\:)$.  Our analytic gap-opening criterion predicts a linear shape of the threshold between this two regions and our SPH simulations allows us to determine the value of it slope. The value that we find for the slope is roughly 0.095.

To increase even more the confidence in our results we test our analytic gap opening criterion against simulations of the literature and we find a good agreement between our gap opening criterion and this simulations.

Our simulations assume only one type of initial conditions for the binary-disk system and we leave for another publication the study of the possible effects of different initial conditions on the results. However, we emphasize that for an isolated binary-disk system the gap opening/closing criterion depends on the local parameters at a given time (e.g. the velocity of the binary-ellipsoid system) but how this local parameters evolves through time depends on the initial conditions, for that reason, our plan in the future is to vary them.  

Finally, we discuss how our results can be applied to the study of the formation of gaps on the circumbinary gas disks around binary MBH and protostars. We discuss the implications of the formation of a gap on the migration and accretion timescales of these systems. Despite the generality of the process that drives the formation of a gap, we remark that is mandatory to explore how more realistic simulations could affect the gap-opening criterion.\\

{\it Acknowledgements}. L del V research was supported by CONICYT Chile (Grant D-21090518) and DAS Universidad de Chile. A.E. acknowledges partial support from the Center of Excellence in Astrophysics and Associated Technologies (PFB 06), FONDECYT Iniciacion Grant 11090216. The simulations were performed using the HPC clusters Markarian (FONDECYT 11090216) and Geryon (PFB 06).

\newpage
\appendix
\section*{APPENDIX A: Initial Conditions}
In our set of simulations, the binary produces a strong non-axisymmetric component to the total gravitational field. Therefore, for the SPH particles near the binary there is not a well defined circular velocity $v(r)$. This is an inherent problem for the initialization of this type of simulations and we chose to solve this by approximating the binary potential by an spherical mass distribution with the same total mass, for computing the initial circular velocity. These spherical representations allow us to compute an initial circular velocity for all the particles of the disk, but the system will not start in a perfect equilibrium.

 We probe different spherical mass distributions to study the effect of the initial velocity on the evolution of the system and then we select the mass distribution that produce the minor overestimation or underestimation of the equilibrium velocity. The spherical mass distributions that we study are: (1) spherical shell (2) point mass distribution and (3) sphere of constant density. The spherical shell (1) produces a discontinuity on the velocity that generates a ring-shaped gap due to an overestimation of the velocity in the region where the gap forms. The point mass distribution (2) overestimates the velocity for all the particles in the region $r<r_{0}$ because in this region the real enclosed mass is less than the mass that is assume by the point mass distribution. This overestimation opens an artificial gap in the center of the disk. The sphere of constant density (3) also generates an overestimation of the velocity in the region $r<r_{0}$ but this overestimation is less intense than the overestimation due to the point mass distribution, because, for the homogeneous sphere, the enclosed mass decreases with radius.

We select the sphere of constant density as the spherical mass distribution for computing the initial velocity, because approximates better the potential of the binary and the system starts closer to a rotationally supported equilibrium. The system is of course not in a perfect equilibrium and our approximation for the initial circular velocity leads to an initial readjustment of the density profile in about an orbital time. For that reason, we only analyze the simulations after 2 initial orbits of the binary, that is, after the system has relaxed into an quasi-equilibrium configuration.

We also include a static Plummer potential to increase the stability of the disk. This static potential decrease the overestimation of the rotational velocity in the inner region of the disk reducing the initial loses of gas in this region and the artificial formation of a gap. In the edges of the disk the cutoff of the disk produces a pressure gradient that drives the expansion of the disk edges. This flow of gas is also reduced due to the gravitational influence of this static potential. We use an static Plummer potential that have a total mass of 12\% the mass of the disk and its core radius is roughly at $0.6\,R_{\rm disk}$. With this selection of parameters the Plummer potential's mass initially enclosed by the orbit of the binary is $\%50$ the total enclosed mass.

\section*{APPENDIX B: Resolution Study}
Our results are derived from the comparison of an analytic criterion for gap-opening against SPH numerical simulations. In the simulations presented in section \S 3, the disk is represented by a collection of $2\times 10 ^5$ SPH particles. In order to check that we have the resolution required and that our results do not depend on that, we run four additional simulations with different numbers of SPH particles; $5\times10^5$, $8\times10^5$ and two with $10^{6}$. 

\begin{figure}[h!] 
\centering
\includegraphics[height=7cm]{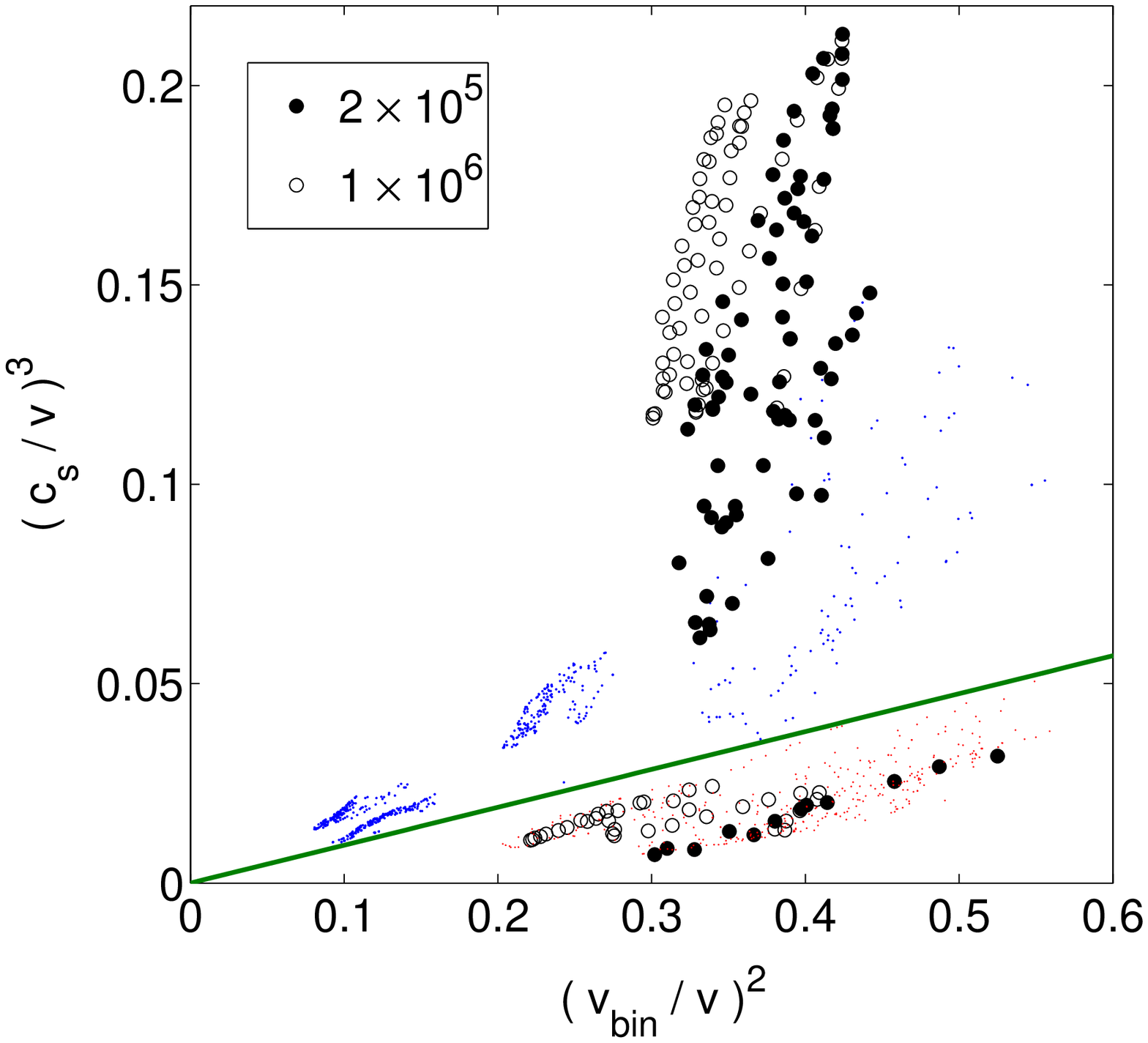}
\caption{ The cubic ratio between the sound speed of the gas and the rotational velocity of the binary-disk system, $ (c_{\rm s}/v)^2$ plotted against the quadratic ratio between the rotational velocity of the isolated binary and the rotational velocity of the binary-disk system $(v_{\rm bin}/v)^2$. The black dots are runs with $2\times10^5$ SPH particles and the black open circles are runs with $1\times10^6$ SPH particles. Also we plot all our others simulations with red ({\it opened} simulations) and blue ({\it closed} simulations) points for comparison. The green line is the interface between this two populations of parameters that is predicted by our analytic gap-opening criterion. The slope of the interface is the function $f(\Delta \phi,\alpha_0 ,\beta_0,\alpha_{\rm ss})$. The two simulations above (below) the green curve with $2\times10^5$ and  $1\times10^6$ SPH particles are simulations with the same initial condition. Although, the evolution of the orbital parameters of the binary are not exactly the same for high and low resolution simulations, the global structure of the disk remain almost unchanged and the region of parameters that the low and high resolution simulations covers is very similar. The classification of these simulations (as {\it closed} or {\it opened}) remain the same for high and low resolution.}
\label{Resolution}
\end{figure}

\begin{figure}[h!] 
\centering
\includegraphics[height=7cm]{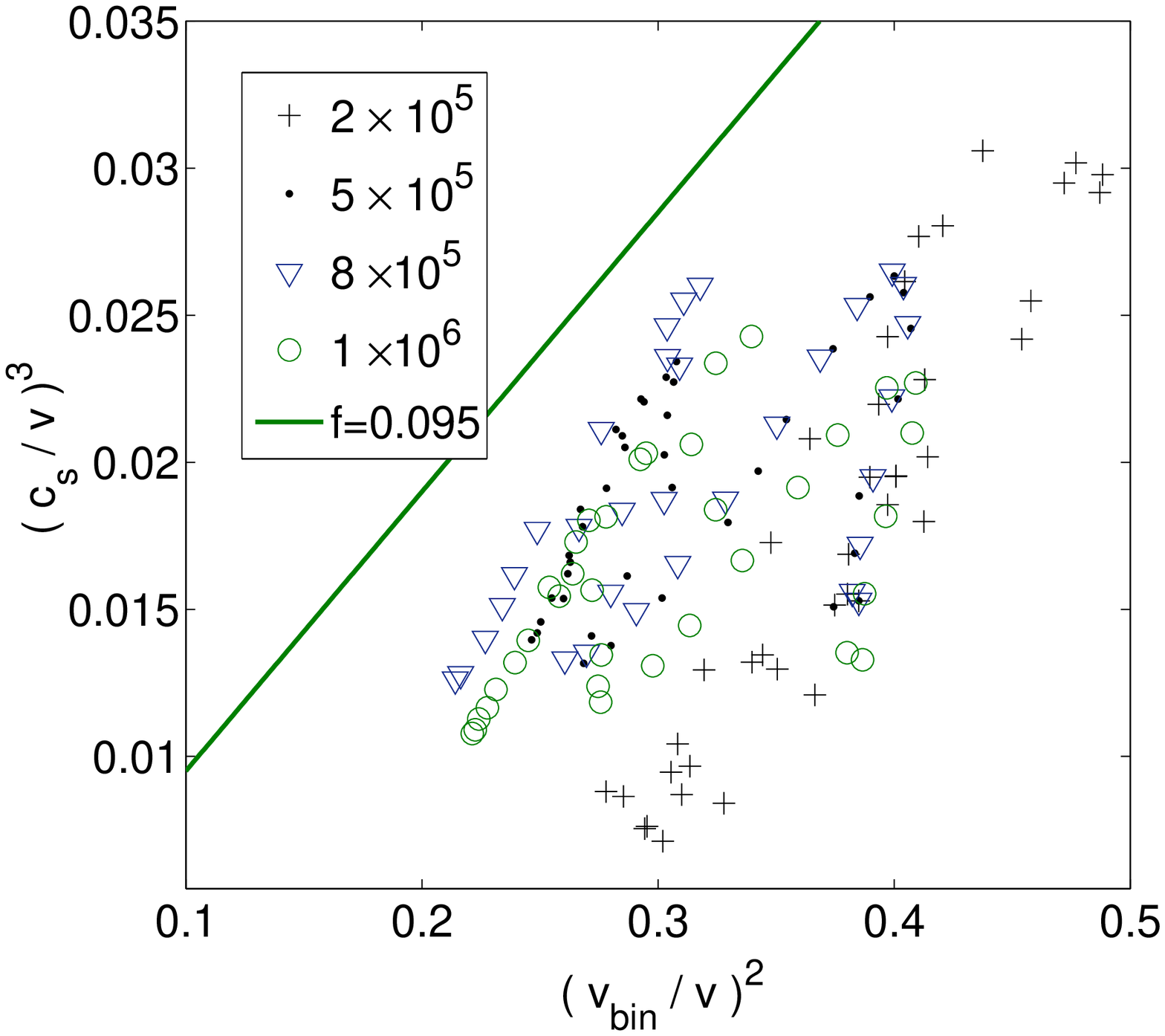}
\caption{ The cubic ratio between the sound speed of the gas and the rotational velocity of the binary-disk system, $ (c_{\rm s}/v)^2$ plotted against the quadratic ratio between the rotational velocity of the isolated binary and the rotational velocity of the binary-disk system $(v_{\rm bin}/v)^2$. In this figure we plot four simulations with the same initial condition but different number of SPH particles: $2\times10^5$, $5\times10^5$, $8\times10^5$ and $1\times10^6$. The green line is the interface between the simulations where a gap is opened in the disk ({\it opened}) and the simulations where the disk dont have a gap ({\it closed}). Although, the evolution of the orbital parameters of the binary are not exactly the same for high and low resolution simulations, the global structure of the disk remain almost unchanged and the region of parameters that the low and high resolution simulations covers is very similar. The classification of these simulations {\it opened} remain the same for high and low resolution.}
\label{Resolution2}
\end{figure}

Although, the evolution of the orbital parameters of the binary are not exactly the same for high and low resolution simulations, the global structure of the disk remain practically almost unchanged. The region of parameters that the low and high resolution simulations covers, in the velocity spaces of parameters, is very similar (figure \ref{Resolution}) and the classification of these simulations (as {\it closed} or {\it opened}) remain the same. For the initial condition that we run with $2\times10^2$, $5\times10^5$, $8\times10^5$ and $10^{6}$ we find that the values of the velocity parameters converge with resolution (figure \ref{Resolution2}). 

\newpage

For the high resolution simulations we find a small shift to higher values of $(c_{\rm s}/v)^3$. This shift can change the value of the slope of the threshold between the {\it closed} and the {\it opened} simulations ($f$) in a $\%10$ or less. Therefore, the overall conclusions for the testing of our analytic gap-opening criterion against SPH simulations remain unchanged. 
\newpage
\section*{References}
\begin{harvard}
\item[] Armitage, P., Natarajan, P. 2002, \apj, {\bf 567L}, 9
\item[] Armitage, P., \& Rice, W. K. M. 2005, {\it Proc. of the STScI May Symp. 2005: A Decade Of Extrasolar Planets Around Normal Stars} p 66 , vol~19, Livio \etal (New York: Cambrige University Press), astro-ph/0507492
\item[] Artymowicz, P., Lubow, S.H. 1994, \apj, {\bf 421}, 651
\item[] Artymowicz, P., Lubow, S.H. 1996, \apj, {\bf 467L}, 77
\item[] Baruteau, C. \& Masset, F. 2012, arXiv:1203.3294
\item[] Barnes, J. E. 2002, \mnras, {\bf 333}, 481
\item[] Barnes, J., Hernquist, L. 1996, \apj, {\bf 471}, 115
\item[] Barnes, J., Hernquist, L. 1992, \araa, {\bf 30}, 705
\item[] Bate, M. R. \& Bonnell, I. A. 1997, \mnras, {\bf 285}, 33
\item[] Bate, M. R., Lubow, S.H., Ogilvie G. I., \& Miller K. A., 2003, \mnras, {\bf 341}, 213
\item[] Begelman, M., Blandford, R., Rees, M.  1980, \nat, {\bf 287}, 307
\item[] Bournaud, F., \etal. 2011, \apj, {\bf 730}, 4
\item[] Boss A. P. 1982 {\it Icarus}, {\bf 51}, 623
\item[] Bryden, G., Chen, X., Lin, D. N. C., Nelson, R. P. \& Papaloizou, J. 1999, \apj, {\bf 514}, 344
\item[] Crida, Morbidelli, \& Masset, 2006, {\it Icarus} {\bf 181}, 587
\item[] Cuadra, J., Armitage, P. J., Alexander, R. D. \&  Begelman, M. 2009, \mnras, {\bf 393}, 1423
\item[] Chapman, S. C., Blain, A. W., Ivison, R. J., \& Smail, Ian R., 2003, {\it Nature}, {\bf 422}, 695  
\item[] Chapman, S. C., Blain, A. W., Smail, Ian R. \& Ivison, R. J., 2005, \apj, {\bf 622}, 772  
\item[] Chapon, D., Mayer, L. \& Teyssier, R. 2011, submitted, eprint arXiv:1110.6086
\item[] Dotti, M., Colpi, M., \& Haardt, F. 2006, \mnras, {\bf 367}, 103
\item[] Downes, D., Solomon, P. M.  1998, \apj, {\bf 507}, 615 (DS)
\item[] Escala, A., del Valle, L., {\it IJMPE}, {\bf 20}, 79 
\item[] Escala, A., Larson, R. B., Coppi, P. S., \& Mardones, D.  2004, \apj, {\bf 607}, 765 
\item[] Escala, A., Larson, R. B., Coppi, P. S., \& Mardones, D. 2005, \apj, {\bf 630}, 152
\item[] Escala, A., 2007 \apj, {\bf 671}, 1264
\item[] Genzel, R., et al. 1998, \apj, {\bf 498}, 579
\item[] Goldreich, P., Tremaine, S. 1980, \apj, {\bf 241}, 425
\item[] Goldreich, P., Tremaine, S. 1982, \araa, {\bf 20}, 249
\item[] Gunther, R., \& Kley, W. 2002, {\it A\&A}, {\bf 387}, 550
\item[] Haiman Z., Kocsis B., Menou K., 2009, \apj, {\bf 700}, 1952
\item[] Kazantzidis, S., \etal 2005, \apj, {\bf 623L}, 67
\item[] Khan F. M., Just A., Merritt D. 2011, \apj, {\bf 732}, 89
\item[] Kocsis, B., Haiman, Z. \& Loeb, A., 2012a, arXiv:1205.4714 
\item[] Kocsis, B., Haiman, Z. \& Loeb, A., 2012b, arXiv:1205.5268
\item[]	Krumholz, M., Klein, R. \& McKee, C., 2007, \apj, {\bf 656}, 959
\item[] Lamb, H. 1879, {\it Hydrodynamics (Cambridge: Cambridge Univ. Press)}
\item[] Lin, D. N. C., Papaloizou, J. 1979, \mnras, {\bf 186}, 799
\item[] Lin, D. N. C., Papaloizou, J. 1986, \apj, {\bf 309}, 846
\item[] Lodato, G., \& Price, D. J. 2010, \mnras, {\bf 405}, 1212
\item[] MacFadyen A. I., Milosavljevic M., 2008, \apj, {\bf 672}, 83
\item[] Mayer, L., Kazantzidis, S., Madau, P., Colpi, M., Quinn, T., \& Wadsley, J. 2007, {\it Science}, {\bf 316}, 1874
\item[] Mayer L., Kazantzidis S., Escala A. \& Callegari S. 2010 \nat, {\bf 466}, 1082
\item[] Mestel, L. 1963, \mnras, {\bf 126}, 553
\item[] Mihos, C., Hernquist, L. 1996, \apj, {\bf 464}, 641
\item[] Milosavljevic, M., Merritt, D. 2001, \apj, {\bf 563}, 34
\item[] Milosavljevic, M., Merritt, D. 2003, \apj, {\bf 596}, 860
\item[] Milosavljevic, M., \& Phinney, E. S. 2005, \apj, {\bf 622}, L93
\item[] Murray, J. R. 1996, \mnras, {\bf 279}, 402
\item[] Nelson, R., \& Papaloizou, J. C. B., \mnras, {\bf 339}, 993
\item[] Plummer, H. C. 1911, \mnras, {\bf 71}, 460
\item[]	Rice, W. K. M. \& Armitage, P. J. 2009, \mnras, {\bf 396} 2228
\item[] Ivanov, P. B., Papaloizou, J. C. B., \& Polnarev, A. G. 1999, \mnras, {\bf 307}, 79
\item[] Richstone, D. \etal 1998, \nat, {\bf 395}, 14
\item[] Roedig, C., Dotti, M., Sesana, A., Cuadra, J. \& Colpi, M. 2011, \mnras {\bf 415} 3033
\item[] Roedig, C., Sesana, A., Dotti, M., Cuadra, J., Amaro-Seoane, P. \& Haardt F., 2012, \mnras, submitted, e-print arXiv:1202.6063
\item[] Sanders, D., Mirabel, I. 1996, \araa, {\bf 34}, 749
\item[] Shakura N. I. \& Sunyaev R. A. 1973, {\it A\&A}, {\bf 24}, 337
\item[] Shi, J., Krolik, J., Lubow, S., \& Hawley, J. 2012, \apj, {\bf 749}, 118
\item[] Springel, V., 2005, \mnras,{\bf 364}, 1105
\item[] Springel, V., Yoshida, N., White, S.  2001, {\it NewA}, {\bf 6}, 79
\item[] Swinbank, A., Smail, I., Longmore, S., Harris, A., Baker, A., De Breuck C., Richard, J., Edge, A., Ivison, R., Blundell, R., Coppin, K., Cox, P., Gurwell, M.,  Hainline, L., Krips, M., Lundgren, A., Neri, R., Siana, B., Siringo, G., Stark, D., Wilner, D., \& Younger, J., 2010 {\it Nature}, {\bf 464}, 733
\item[] Taylor, P. \& Miller J., 2011, arXiv:1112.5120v3 [astro-ph.IM]
\item[] Takeuchi, T. \& Lin, D. N. C., 2002, \apj, {\bf 581}, 1344
\item[] Takoni, L., Neri, R., Chapman S., Genzel R., Smali, I., Ivison, R., Bertoldi, F,. Blin A., Cox P., Greve T. \& Omont A., 2006, \apj, {\bf 640}, 228
\item[] Taniguchi, Y., Wada, K. 1996, \apj, {\bf 469}, 581
\item[] Ward, W. R, \& Hourigan, K. 1989, \apj, {\bf 347}, 490
\item[] Ward, W. R. 1997, {\it Icarus}, {\bf 126}, 261

\end{harvard}


\end{document}